# Celeste: Variational inference for a generative model of astronomical images

**Jeffrey Regier,** University of California, Berkeley        JEFF@STAT.BERKELEY.EDU
**Andrew Miller,** Harvard University        ACM@SEAS.HARVARD.EDU
**Jon McAuliffe,** University of California, Berkeley        JON@STAT.BERKELEY.EDU
**Ryan Adams,** Harvard University        RPA@SEAS.HARVARD.EDU
**Matt Hoffman,** Adobe Research        MDHOFFMA@CS.PRINCETON.EDU
**Dustin Lang,** Carnegie Mellon University        DSTN@CMU.EDU
**David Schlegel,** Lawrence Berkeley National Laboratory        DJSCHLEGEL@LBL.GOV
**Prabhat,** Lawrence Berkeley National Laboratory        PRABHAT@LBL.GOV

## Abstract

We present a new, fully generative model of optical telescope image sets, along with a variational procedure for inference. Each pixel intensity is treated as a Poisson random variable, with a rate parameter dependent on latent properties of stars and galaxies. Key latent properties are themselves random, with scientific prior distributions constructed from large ancillary data sets. We check our approach on synthetic images. We also run it on images from a major sky survey, where it exceeds the performance of the current state-of-the-art method for locating celestial bodies and measuring their colors.

## 1. Introduction

This paper presents Celeste, a new, fully generative model of astronomical image sets—the first such model to be empirically investigated, to our knowledge. The work we report is an encouraging example of principled statistical inference applied successfully to a science domain underserved by the machine learning community. It is unfortunate that astronomy and cosmology receive comparatively little of our attention: the scientific questions are fundamental, there are petabytes of data available, and we as a data-analysis community have a lot to offer the domain scientists. One goal in reporting this work is to raise the profile of these problems for the machine-learning audience and show that much interesting research remains to be done.

Turn now to the science. Stars and galaxies radiate photons. An astronomical image records photons—each originating from a particular celestial body or from background atmospheric noise—that pass through a telescope's lens during an exposure. Multiple celestial bodies may contribute photons to a single image (e.g. Figure 1), and even to a single pixel of an image. Locating and characterizing the imaged celestial bodies is an inference problem central to astronomy. To date, the algorithms proposed for this inference problem have been primarily heuristic, based on finding bright regions in the images (Lupton et al., 2001; Stoughton et al., 2002).

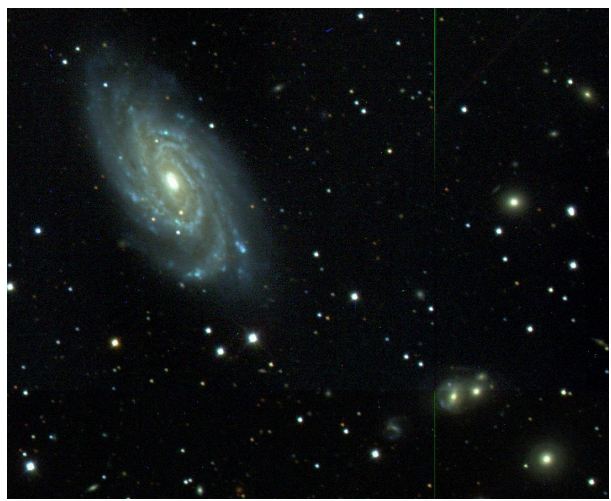

Figure 1. An image from the Sloan Digital Sky Survey (SDSS, 2015) of a galaxy from the constellation Serpens, 100 million light years from Earth, along with several other galaxies and many stars from our own galaxy.

Generative models are well-suited to this problem—for three reasons. First, to a good approximation, photon counts from celestial objects are independent Poisson processes: each star or galaxy has an intrinsic brightness that is effectively static during human time scales. In an imaging exposure, the expected count of photons entering the telescope's lens from a particular object is proportional to its brightness. When multiple objects contribute photons to the same pixel, their rates combine additively.

Second, many sources of prior information about celestial bodies are available, but none is definitive. Stars tend to be brighter than galaxies, but many stars are dim and many galaxies are bright. Stars tend to be smaller than galaxies, but many galaxies appear point-like as well. Stars and galaxies differ greatly in how their radiation is distributed over the visible spectrum: stars are well approximated by an "ideal blackbody law" depending only on their temperature, while galaxies are not. On the other hand, stars are not actually ideal blackbodies, and galaxies do emit energy in the same wavelengths as stars. Posterior inference in a generative model provides a principled way to integrate these various sources of prior information.

Third, even the most powerful telescopes receive just a handful of photons per exposure from many celestial objects. Hence, many objects cannot be precisely located, classified, or otherwise characterized from the data available. Quantifying the uncertainty of point estimates is essential—it is often as important as the accuracy of the point estimates themselves. Uncertainty quantification is a natural strength of the generative modeling framework.

Some astronomical software uses probabilities in a heuristic fashion (Bertin & Arnouts, 1996), and a generative model has been developed for measuring galaxy shapes (Miller et al., 2013)—a subproblem of ours. But, to our knowledge, fully generative models for inferring celestial bodies' locations and characteristics have not yet been examined.[1] Difficulty scaling the inference for expressive generative models may have hampered their development, as astronomical sky surveys produce very large amounts of data. For example, the Dark Energy Survey's 570-megapixel digital camera, mounted on a four-meter telescope in the Andes, captures 300 gigabytes of sky images every night (Dark Energy Survey, 2015). Once completed, the Large Synoptic Survey Telescope will house a 3200-megapixel camera producing eight terabytes of images nightly (Large Synoptic Survey Telescope Consortium, 2014).

The rest of the paper describes the Celeste model (Section 2) and its accompanying variational inference procedure (Section 3). Section 4 details our empirical studies on synthetic data as well as a sizable collection of astronomical images.

---

[1]However, see Hogg (2012) for a workshop presentation proposing such a model.

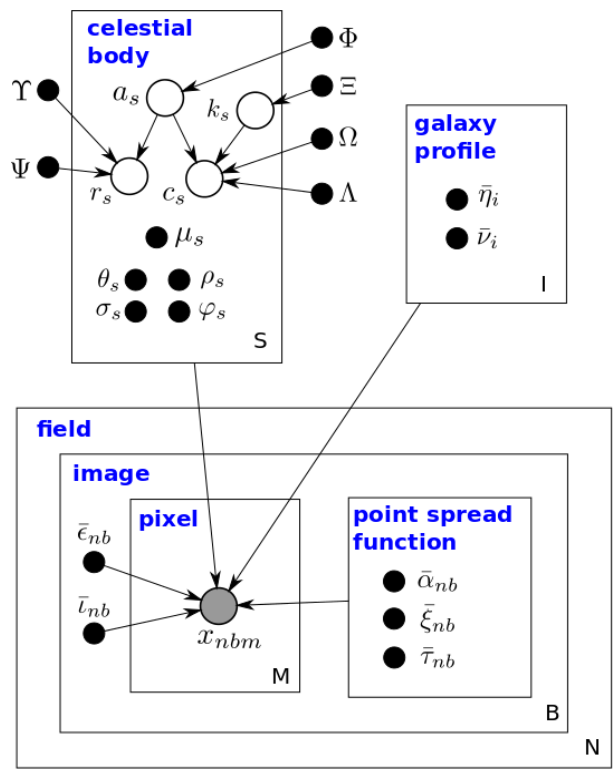

Figure 2. The Celeste graphical model. Shaded vertices represent observed random variables. Empty vertices represent latent random variables. Black dots represent constants. Constants with "bar" decorators, e.g. $\bar{\epsilon}_{nb}$, are set a priori. Constants denoted by uppercase Greek characters are also fixed; they denote parameters of prior distributions. The remaining constants and all latent random variables are inferred. Edges signify conditional dependency. Rectangles ("plates") represent independent replication.

## 2. The model

The Celeste model is represented graphically in Figure 2. In this section we describe how Celeste relates celestial bodies' latent characteristics to the observed pixel intensities in each image.

### 2.1. Celestial bodies

Celeste is a hierarchical model, with celestial objects atop pixels. For each object $s = 1, \ldots, S$, the unknown 2-vector $\mu_s$ encodes its position in the sky as seen from earth. In Celeste, every celestial body is either a star or a galaxy. (In the present work, we ignore other types of objects, which are comparatively rare.) The latent Bernoulli random variable $a_s$ encodes object type: $a_s = 1$ for a galaxy, $a_s = 0$ for a star. We set the prior distribution

$$a_s \sim \text{Bernoulli}(\Phi). \qquad (1)$$

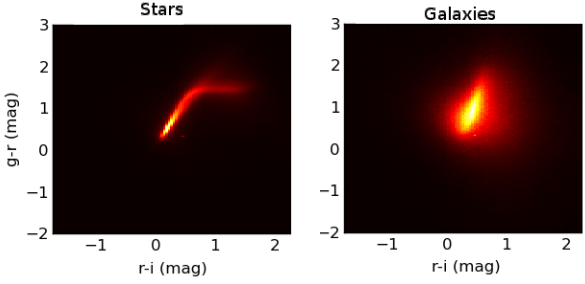

*Figure 3.* Density plots for two colors, based on an SDSS catalog containing hundreds of thousands of stars and galaxies.

### 2.1.1. BRIGHTNESS

The overall brightness of a celestial object $s$ is quantified as the total radiation from $s$ expected to reach a unit area of the earth's surface, directly facing $s$, per unit of time. However, we can also quantify brightness as the proportion of this radiation (per square meter, per second) that passes through each filter in a standardized filter set. Such a set is called a "photometric system." These standardized filters are approximately band-pass: each allows most of the energy in a certain band of wavelengths through, while blocking most of the energy outside the band. The physical filters attached to a telescope lens closely match the standardized filters of some photometric system.

In Celeste, we take the photometric-system approach—we directly model brightnesses with respect to the $B$ filters of a fixed photometric system. We designate a particular filter as the "reference" filter, letting the random variable $r_s$ denote the brightness of object $s$ with respect to that filter. We make $r_s$ dependent on $a_s$, since stars tend to be brighter than galaxies. For computational convenience, and because brightness is typically considered to be non-negative and real-valued, we set

$$r_s | (a_s = i) \sim \text{Gamma}\left(\Upsilon^{(i)}, \Psi^{(i)}\right). \qquad (2)$$

Object $s$'s brightnesses with respect to the remaining $B-1$ filters are encoded using "colors." The color $c_{sb}$ is defined as the log ratio of brightnesses with respect to filters $b$ and $b+1$. Here, the filters are ordered by the wavelength bands they let through. The $B-1$ colors for object $s$ are collectively denoted by $c_s$, a random $(B-1)$-vector.

Celeste uses the color parameterization because stars and galaxies have very distinct prior distributions in color space. Indeed, for idealized stars—blackbodies—all $B-1$ colors lie on a one-dimensional manifold indexed by surface temperature. On the other hand, though galaxies are composed of stars, theory does not suggest they lie near the same manifold: the stars in a galaxy can have many different surface temperatures, and some of the photons are re-processed to other energies through interactions with dust and gas. (Figure 3) demonstrates that stars are much nearer a 1-dimensional manifold in color space than galaxies are.

We model the prior distribution on $c_s$ as a mixture of $D$ multivariate Gaussians. The number of mixture components $D$ may be selected to minimize error on held out data, or kept small for computational efficiency. The random categorical variable $k_s$ indicates which mixture component generated $c_s$. A priori,

$$k_s | (a_s = i) \sim \text{Categorical}(\Xi_1, \ldots, \Xi_D) \qquad (3)$$

and

$$c_s | (k_s = d, a_s = i) \sim \text{MvNormal}\left(\Omega^{(i,d)}, \Lambda^{(i,d)}\right). \quad (4)$$

A celestial body's brightnesses in the $B$ filters, $(\ell_{sb})_{b=1}^{B}$, is uniquely specified by its reference-filter brightness $r_s$ and its colors $c_s$.

### 2.1.2. GALAXIES

The distance from earth to any star (besides the sun) exceeds the star's radius by many orders of magnitude. Therefore, stars are well modeled as points. Modeling the (two-dimensional) appearance of galaxies as seen from earth is more involved. We divide the appearance of galaxy $s$ into two parts: its per-filter brightnesses ($\ell_{sb}$), as discussed in Section 2.1.1, and its "light kernel" $h_s(w)$, which describes how the apparent radiation from the galaxy is distributed over the sky. The argument $w$ is in sky coordinates; the light kernel is a density function that integrates to one and is largest near the apparent galactic center. In Section 2.1.3, we show how these two parts of apparent galaxy brightness come together.

We take $h_s(w)$ to be a convex combination of two prototype functions, known in astronomy as the "exponential" and "de Vaucouleurs" prototypes:

$$h_s(w) = \theta_s h_{s1}(w) + (1 - \theta_s) h_{s2}(w), \quad \theta_s \in [0, 1]. \quad (5)$$

The de Vaucouleurs prototype is characteristic of "elliptical" galaxies, which have smooth light kernels (Figure 5), whereas the exponential prototype matches "spiral" galaxies (Figure 1). The prototype functions $h_{s1}(w)$ and $h_{s2}(w)$ contain additional galaxy-specific parameters. In particular, each prototype function is a rotated, scaled mixture of bivariate normal distributions. The rotation and scaling are galaxy-specific, but the remaining parameters of each mixture are not:

$$h_{si}(w) = \sum_{j=1}^{J} \bar{\eta}_{ij} \phi(w; \mu_s, \bar{v}_{ij} W_s), \quad i = 1 \text{ or } 2. \quad (6)$$

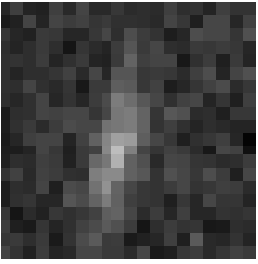

*Figure 4.* A distant galaxy, from the SDSS dataset, approximately 20 pixels in height, predicted to have effective radius $\sigma_s = 0.6$ arcseconds, rotation $\varphi_s = 80°$, and eccentricity $\rho_s = 0.17$. Credit: SDSS

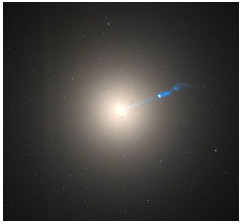

*Figure 5.* Messier 87, a galaxy that exhibits the de Vaucouleurs profile. Credit: NASA

In Equation (6), $(\bar{\eta}, \bar{v})_{ij}$ are pre-specified constants that characterize the exponential and de Vaucouleurs prototypes; $\mu_s$ is the center of the galaxy, in sky coordinates; $W_s$ is a spatial covariance matrix; and $\phi$ is the bivariate normal density.

The light kernel $h_s(w)$ is a finite scale mixture of Gaussians: its mixture components have a common mean $\mu_s$ and covariance matrices that differ only in scale. The "isophotes" (level sets of the light kernel) are concentric ellipses. Although this model prevents us from fitting individual "arms," like those of the galaxy in Figure 1, most galaxies are not sufficiently resolved to see such substructure. (See Figure 4 for a typical galaxy image.) A more flexible galaxy model might overfit them.

The spatial covariance matrix $W_s$ is parameterized by a rotation angle $\varphi_s$, an eccentricity (minor-major axis ratio) $\rho_s$, and an overall size scale $\sigma_s$:

$$W_s = R_s^{\top} \begin{bmatrix} \sigma_s^2 & 0 \\ 0 & \sigma_s^2 \rho_s^2 \end{bmatrix} R_s, \qquad (7)$$

where $R_s$ is a rotation matrix,

$$R_s = \begin{bmatrix} \cos \varphi_s & -\sin \varphi_s \\ \sin \varphi_s & \cos \varphi_s \end{bmatrix}. \qquad (8)$$

The scale $\sigma_s$ is specified in terms of "effective radius"—the radius of the disc that contains half of the galaxy's light emissions before applying the eccentricity $\rho_s$.

### 2.1.3. THE IDEAL SKY VIEW

In the upcoming Section 2.2, we account for distortions from pixelation, atmospheric blur, and background noise; and we deal in photons counted by a camera, rather than continuous-valued brightness (energy arriving at earth).

The developments of the previous sections represent the sky without these concerns; we call this part of the model the "ideal sky view." Let $\delta_{\mu_s}$ denote the Dirac delta function—the light kernel of a star. Then the total apparent brightness (the ideal sky view) in filter $b$, at sky position $w$, is

$$G_b(w) = \sum_{s=1}^{S} \ell_{sb} g_{sa_s}(w) \qquad (9)$$

where

$$g_{si}(w) = \begin{cases} \delta_{\mu_s}(w), & \text{if } i = 0 \text{ ("star")} \\ h_s(w), & \text{if } i = 1 \text{ ("galaxy")}. \end{cases} \qquad (10)$$

## 2.2. Astronomical images

Returning to the Celeste graphical model, the data is represented in the bottom half of Figure 2. A "field" is a small, rectangular region of the sky. Fields may overlap. Each of the $N$ fields in the data set is imaged $B$ times, once per filter in the photometric system (Section 2.1.1).[2]

Each resulting image is a grid of $M$ pixels. Each pixel in turn receives light primarily from celestial bodies near the pixel's corresponding region of the sky. The observed random variable $x_{nbm}$ is the count of photons recorded at pixel $m$, in image $b$ of field $n$.

The night sky is not completely dark owing to natural skyglow, a combination of reflected sunlight off dust particles in the solar system, night airglow from molecules in the earth's atmosphere, and scattered starlight and moonlight. We model the night sky's brightness as background noise, through a spatial Poisson process that is homogeneous for each image and independent of stars and galaxies. The noise rate depends on the image and the band because atmospheric conditions vary over time. Also, the atmospheric effects differ for imaging targets closer to the horizon. The image-specific constant $\bar{\epsilon}_{nb}$ denotes the rate of background noise.

Both $\bar{\epsilon}_{nb}$ and the brightnesses of celestial bodies are quantified in linear units of nanomaggies—a physical unit of energy, not specific to any particular image (SDSS DR10, 2015). The image-specific constant $\bar{\iota}_{nb}$ is the expected number of photons recorded in image $b$ of field $n$, for a pixel receiving a brightness of one nanomaggy.

### 2.2.1. POINT-SPREAD FUNCTIONS

Ground-based astronomical images are blurred by a combination of small-angle scattering in the earth's atmosphere,

---

[2] Cameras for optical astronomy typically use charge-coupled devices (CCDs), which convert light into electrons (Dark Energy Survey, 2015). A CCD is a grid of millions of pixels, designed to sum the radiation arriving at each pixel during an exposure.

the diffraction limit of the telescope, optical distortions in the camera, and charge diffusion within the silicon of the CCD detectors. Together these effects are represented by the "point spread function" (PSF) of a given image. Stars (other than the sun) are points in the ideal sky view (Section 2.1.3), but the PSF typically spreads their photons over dozens of adjacent pixels.

When dealing with images, as in this section, we work in the image coordinate system. For any single image, there is a one-to-one mapping between image and sky coordinates, so nothing is lost; our notation suppresses the mapping for clarity.

We model the action of the PSF as a mixture of $K$ Gaussians. Consider pixel $m$ (in band $b$ of image $n$), having coordinates $w_m$. The probability that a photon originating at coordinates $w$ lands at $w_m$ is given by the PSF

$$f_{nbm}(w) = \sum_{k=1}^{K} \bar{\alpha}_{nbk} \phi(w_m; w + \bar{\xi}_{nbk}, \bar{\tau}_{nbk}). \quad (11)$$

Here $\phi$ is the bivariate normal density. The parameters $(\bar{\alpha}_{nb}, \bar{\xi}_{nb}, \bar{\tau}_{nb})$ of the the PSF are specific to an image, in part to account for atmospheric conditions that vary between exposures, but are constant throughout the image.

### 2.2.2. The Celeste likelihood

Convolving the ideal sky view (Equation 9) with the PSF and adding background noise yields the rate of photon arrivals for pixel $m$:

$$z_{nbm} := \epsilon_{nb} + \int G_b(w) f_{nbm}(w) dw \quad (12)$$

$$= \epsilon_{nb} + \sum_{s=1}^{S} \ell_{sb} \int g_{sas}(w) f_{nbm}(w) dw. \quad (13)$$

These normal-normal convolutions are analytic. For stars ($a_s = 0$),

$$\check{f}_{s0}(m) := \int g_{s0}(w) f_{nbm}(w) dw \quad (14)$$

$$= \sum_{k=1}^{K} \bar{\alpha}_{nbk} \phi(w_m; \mu_s + \bar{\xi}_{nbk}, \bar{\tau}_{nbk}). \quad (15)$$

Let $\theta_{s1} = \theta_s$ and $\theta_{s2} = 1 - \theta_s$. For galaxies ($a_s = 1$),

$$\check{f}_{s1}(m) := \int g_{s1}(w) f_{nbm}(w) dw \quad (16)$$

$$= \sum_{k=1}^{K} \bar{\alpha}_{nbk} \sum_{i=1}^{2} \theta_{si} \sum_{j=1}^{J} \bar{\eta}_{ij}$$
$$\cdot \phi(w_m; \mu_s + \bar{\xi}_{nbk}, \bar{\tau}_{nbk} + \bar{v}_{ij} W_s). \quad (17)$$

Let $a = (a_s)_{s=1}^{S}$, $r = (r_s)_{s=1}^{S}$, and $c = (c_s)_{s=1}^{S}$. Then the expected number of photons landing in pixel $m$ is

$$F_{nbm}(a, r, c) = \bar{\iota}_{nb}[\epsilon_{nb} + z_{nbm}]. \quad (18)$$

For $n = 1, \ldots, N$, $b = 1, \ldots, B$, and $m = 1, \ldots, M$, we finally obtain the likelihood

$$x_{nbm} | a, r, c \stackrel{\text{ind}}{\sim} \text{Poisson}(F_{nbm}(a, r, c)). \quad (19)$$

## 3. Inference

In this section we explain how we apply the Celeste model to astronomical image data sets, to draw inferences about unknown quantities.

In principle, all parameters could be learned by variational inference. But we reuse some estimates from the existing photometric pipeline that are not thought to limit performance. The background noise level $\bar{\epsilon}_{bn}$ is set by the existing photometric pipeline, based on a heuristic: most pixels in each image receive photons primarily from background radiation. The calibration constant $\bar{\iota}_{bn}$ is set by first calibrating overlapping images relative to each other, and then by calibrating some images absolutely, based on benchmark stars (Padmanabhan et al., 2008). The image-specific parameters of the point spread function, $(\bar{\alpha}_{nb}, \bar{\eta}_{nb}, \bar{\tau}_{nb})_{k=1}^{K}$, are set by the existing photometric pipeline; a mixture of Gaussians is fit to known stars (considered point sources, before convolution with the PSF) in each image that are not near other celestial objects.

Galaxy profiles' parameters, $(\bar{\eta}_{ij}, \bar{v}_{ij})_{j=1}^{J}$, are set a priori too, by fitting mixtures of Gaussians to Sérsic profiles—widely used models of galaxy profiles. We fit $J = 8$-component mixtures for de Vaucouleurs galaxies ($i = 1$) and $J = 6$-component mixtures for exponential galaxies ($i = 2$). These approximations are good enough for the current version of the model: the misfit versus Sérsic profiles is smaller than the misfit between Sérsic profiles and actual galaxies.

The prior distributions also could be learned within the variational procedure, by empirical Bayes. But, because much data is available in existing astronomical catalogs, we estimate their parameters a priori. Fitting $\Phi$, $\Upsilon$, and $\Lambda$ by maximum likelihood to existing catalogs is straightforward. To fit the prior on color ($c_s$) to existing catalogs, we use the expectation-maximization algorithm, initialized by k-means (van Leeuwen, 2015). Though $D = 64$ minimized held-out test error, we set $D = 2$, to work around a limitation of our present optimizer—it only supports box constraints. Figure 3 shows a two-dimensional slice of the $(B - 1)$-dimensional data set used to construct the color prior.

The remaining quantities are estimated by variational inference.

### 3.1. Variational approximation of the posterior distribution

Let $\Theta = (a, r, k, c)$ be the latent random variables in the Celeste model, and let $x = (x_{111}, \ldots, x_{NBM})$ be the pixel intensities. Computing the posterior $p(\Theta|x)$ is intractable: to apply Bayes's rule exactly, we need to evaluate

$$p(x) = \int \prod_{n=1}^{N} \prod_{b=1}^{B} \prod_{m=1}^{M} p(x_{nbm}|\Theta) \prod_{s=1}^{S} p(\Theta_s) \, d\Theta. \quad (20)$$

But the $S$-dimensional integral in (20) does not decompose into a product of low-dimensional integrals, and therefore cannot be computed numerically—nor does it have a closed form.

Instead, we use optimization to find a distribution that best approximates the posterior. For any distribution $q$ over $\Theta$,

$$\log p(x) \geq \mathbb{E}_q[\log p(x, \Theta)] - \mathbb{E}_q[\log q(\Theta)] \quad (21)$$

$$=: \mathscr{L}(q). \quad (22)$$

Here $\mathbb{E}_q$ is expectation with respect to $q$. We call $\mathscr{L}$ the evidence lower bound (ELBO). To find a distribution $q^{\star}$ that approximates the exact posterior, we maximize the ELBO over a set $\mathcal{Q}$ of candidate $q$'s. We restrict $\mathcal{Q}$ to distributions of the factored form

$$q(\Theta) = \prod_{s=1}^{S} q(a_s) q(r_s|a_s) q(k_s|a_s) \prod_{b=1}^{B-1} q(c_{sb}|a_s). \quad (23)$$

Furthermore, for all $s = 1, \ldots, S$, $b = 1, \ldots, B-1$, and $i \in \{0, 1\}$, we set

$$q(a_s) \sim \text{Bernoulli}(\chi_s), \quad (24)$$

$$q(r_s|a_s = i) \sim \text{Gamma}(\gamma_s^{(i)}, \zeta_s^{(i)}), \quad (25)$$

$$q(k_s|a_s = i) \sim \text{Categorical}(\kappa_s^{(i)}), \quad (26)$$

$$q(c_{sb}|a_s = i) \sim \text{Normal}(\beta_{sb}^{(i)}, \lambda_{sb}^{(i)}). \quad (27)$$

Then finding $q^{\star}$ is equivalent to finding the optimal $(\chi_s, \gamma_s, \zeta_s, \kappa_s, \beta_s, \lambda_s)$ for each celestial body. By design, most of the expectations in $\mathscr{L}$ can be evaluated analytically. From the density function for the Poisson distribution,

$$\mathbb{E}_q[\log p(x|\Theta)] \quad (28)$$

$$= \sum_{n=1}^{N} \sum_{b=1}^{B} \sum_{m=1}^{M} \Big\{ x_{nbm} \mathbb{E}_q[\log F_{nbm}] \\ - \mathbb{E}_q[F_{nbm}] - \log(x_{nbm}!) \Big\}. \quad (29)$$

Space constraints prevent us from showing the derivation of analytic expectations here.

One expectation is not analytic. We approximate it by replacing a logarithm with its second-order Taylor expansion:

$$\mathbb{E}_q[\log z_{nbm}] \approx \log \mathbb{E}_q[z_{nbm}] - \frac{\mathbb{V}_q[z_{nbm}]}{2\{\mathbb{E}_q[z_{nbm}]\}^2}, \quad (30)$$

where $\mathbb{V}_q$ denotes variance with respect to $q$. This technique is known as delta-method variational inference (Braun & McAuliffe, 2010; Wang & Blei, 2013). Because $z_{nbm}$ is a sum over celestial bodies, whose corresponding random variables are treated as independent in $q$, the computational complexity of the approximation scales linearly in the number of celestial bodies:

$$\mathbb{V}_q[z_{nbm}] = \sum_{s=1}^{S} \mathbb{V}_q[\ell_{sb} \breve{f}_{sa_s}(m)]. \quad (31)$$

### 3.2. Optimization

Once the expectations in the ELBO are replaced with analytic expressions, maximizing it becomes a standard optimization problem, amenable to various techniques. We use L-BFGS-B (Byrd et al., 1995).

When possible, we use existing star and galaxy catalogs for initialization. When no catalog is suitable, we convolve the images with matched filters to increase the signal-to-noise ratio (North, 1963). We find each pixel whose value exceeds both the values of its neighboring pixels and an upper bound on the number of photons that could come from sky noise. We initialize the center of each such pixel as a celestial object. The number of such pixels determines $S$, the number of objects we assume are present in the image. Modeling $S$ as random is the subject of ongoing research.

We compute derivatives alongside the evaluation of the objective, with little overhead: results for the most computationally expensive operations required to evaluate the objective, like exponentiation, can be reused for evaluating the derivative. We have not found the speed of automatic differentiation toolkits competitive with manually coded derivatives, so our current results use the latter. Validating manually derived derivatives against approximations from numeric differentiation is essential.

To get a feel for the scale of the computation, consider how we produced the results in the next section. The objective and its derivative are summations over pixels (Equation (29)). At each pixel, for each nearby celestial body, we evaluate 45 Gaussian densities, to compute the quantities in Equations (15) and (17). In the model, every celestial body can contribute photons to every pixel. In practice, we truncate these Gaussians. Most stars and galaxies contribute photons to fewer than 100 pixels, and no pixels are thought to receive photons from more than 10 celestial bodies.

With these techniques, evaluating the ELBO takes several seconds on a 2000-pixel image containing a few celestial bodies, and roughly 5 minutes on a 4-megapixel image with hundreds of celestial bodies. The calculation for a single image could be parallelized, though instead we elect to process images in parallel, on separate processors.

## 4. Experiments

For real astronomical images, ground truth is unknown. However, a region of the sky known as "Stripe 82" has been imaged more than 30 times by modern telescopes, whereas most of the sky has been imaged through all five filter bands of our chosen photometric system just once. "Photo" (Lupton et al., 2001; Lupton, 2005) is the current state of the art for detection and characterization of celestial bodies. We henceforth refer to Photo, limited to just one image in each band, as "Primary," and Photo run on the complete collection of replicated Stripe 82 images as "Coadd." Coadd serves as the ground truth in our subsequent analysis. However: (1) any systematic biases in the Photo software are shared by both Coadd and Primary, and (2) the composition of images taken through different atmospheric conditions can create its own biases. Nonetheless, with predictions based on at least 30 times more data, we expect that Coadd accurately characterizes any celestial body detected by either Primary or Celeste.

We compare Celeste to Primary on 654 celestial bodies from Stripe 82, selected based on Coadd (the ground truth), that were not so bright as to be trivial to detect, but not so dim as to be impossible to detect. The data are sets of $B = 5$ images, called "stamps", each centered on a selected celestial body. For these experiments, stamps substitute for fields in Figure 2. Unlike fields, the stamps do not overlap, and a celestial body appearing in one stamp does not contribute photons to other stamps that we analyze. Multiple celestial bodies contribute photons to most stamps.

We initialize Celeste to the output of Primary, so that we can assess the marginal improvement obtained from our inference procedure. Results appear in Table 1.

Sample size varies by row because the galaxy models for Celeste and Photo are not always comparable: both fit exponential and de Vaucouleurs light-kernel prototypes to each galaxy, but in Celeste both prototypes are constrained to have the same rotation and scaling applied to them. Hence, for rotation and scaling measures, we only compare using galaxies where Coadd puts all the mixing weight on one of the two prototypes.

Primary (Photo) is a carefully hand-tuned heuristic. Yet, Celeste matches or improves Photo on most metrics; only for reference-band brightness and scale is Celeste worse by more than two standard errors. This result comes on a data set of difficult celestial bodies, with ground truth set by Photo itself. For each color, Celeste reduces Primary's error, making Celeste (initialized by Photo) state-of-the-art for color detection. Whereas Primary estimates each filter band's brightness independently, Celeste predicts band brightnesses jointly, and regularizes these predictions based on prior information.

Celeste significantly outperforms Primary for position too. The 9.8% (+/- 2.0%) smaller position error is of practical importance to astronomers.

### 4.1. Synthetic images

To gauge the extent to which Celeste's performance is limited by model misfit, or by errors in Coadd (the ground truth), we also test Celeste with synthetic images. For each real image, we generate a synthetic image with the same properties, with the locations, celestial object types, and reference band bright of the celestial bodies set to Coadd, but with the colors and pixel values drawn from our model. For each synthetic image, we initialize Celeste to the predictions from Primary (run on the real images, not the synthetic images).

We would like to also test Photo on the synthetic images. Running Photo on new data, however, rather than using the catalogs from former runs of Photo, exceeds our capacity; Photo is a long, intricate hand-tuned pipeline that has not been compiled in 6 years. Comparing the results for Celeste on synthetic data to Primary is nonetheless informative, since the synthetic data mirrors Coadd (the ground truth).

For position, brightness, color, and all 4 properties of galaxies, we see large reductions in error by Celeste from using synthetic images rather than real images. The number of galaxies we misclassify as stars, and vice versa, is also reduced.

### 4.2. Uncertainty quantification

Celeste is fairly certain ($\leq 1\%$ uncertainty) about the classification (star vs. galaxy) for 526 out of 573 celestial objects. Of these classifications, 4% are wrong. Of the remaining classifications ($>1\%$ uncertainty), 45% are wrong.

For brightness and each of the four colors, observed error rates also correlate directly with reported uncertainty (Table 2). This correlation holds for synthetic data too.

*Table 1.* Columns 1 and 2 are the average error for Primary and Celeste on celestial bodies from Stripe 82; column 4 is the average error for Celeste on synthetic images. **Lower is better.** "Improve" is the improvement from Celeste relative to Primary, with SE in parentheses. Celeste on synthetic data is compared to Primary on real data; see the text. "N" is sample size. "Position" is error, in pixels, for the location of the celestial bodies' centers. "Missed gals" counts galaxies labeled as stars. "Missed stars" counts stars labeled as galaxies. "Brightness" measures the reference band (r-band) brightness in nanomaggies. "Colors" are log-ratios of brightnesses in consecutive bands. "Profile" is a proportion indicating whether a galaxy is de Vaucouleurs or exponential. "Eccentricity" is the ratio between the lengths of a galaxy's minor and major axes. "Scale" is the effective radius of a galaxy in arcsecs. "Angle" is the orientation of a galaxy in degrees.

| DATASET | REAL | | | SYNTHETIC | | |
|---|---|---|---|---|---|---|
| MODEL | PRIMARY | CELESTE | IMPROVE | CELESTE | IMPROVE | N |
| POSITION | 0.22 | **0.20** | .02 (.00) | 0.08 | .14 (.01) | 654 |
| MISSED GALS | 28 / 654 | **15 / 654** | .02 (.01) | 14 / 654 | .02 (.01) | 654 |
| MISSED STARS | **8 / 654** | 31 / 654 | -.04 (.01) | 6 / 654 | .00 (.01) | 654 |
| BRIGHTNESS | **0.76** | 1.60 | -.83 (.12) | 0.29 | .47 (.08) | 654 |
| COLOR U-G | 1.10 | **0.49** | .61 (.04) | 0.20 | 1.06 (.05) | 582 |
| COLOR G-R | 0.16 | **0.09** | .07 (.01) | 0.05 | .43 (.02) | 654 |
| COLOR R-I | 0.09 | **0.06** | .03 (.00) | 0.04 | .25 (.01) | 654 |
| COLOR I-Z | 0.25 | **0.10** | .15 (.01) | 0.08 | .31 (.02) | 654 |
| PROFILE | 0.19 | 0.23 | -.04 (.02) | 0.16 | .03 (.02) | 237 |
| ECCENTRICITY | 0.17 | **0.13** | .04 (.01) | 0.11 | .05 (.01) | 237 |
| SCALE | **0.37** | 1.28 | -.91 (.17) | 0.23 | .14 (.04) | 237 |
| ANGLE | 19.40 | 18.10 | 1.40 (.80) | 14.90 | 4.50 (.80) | 237 |

*Table 2.* Average error for Celeste's predictions, for real astronomical images, binned into quartiles by estimated uncertainty. SEs in parentheses. Bins of more uncertain predictions have greater average error, without exception.

| | Q1 | Q2 | Q3 | Q4 |
|---|---|---|---|---|
| BRIGHTNESS | .27 (.02) | .53 (.04) | .94 (.10) | 4.04 (.50) |
| COLOR U-G | .17 (.01) | .43 (.04) | .65 (.07) | .85 (.09) |
| COLOR G-R | .05 (.00) | .07 (.01) | .09 (.01) | .15 (.01) |
| COLOR R-I | .03 (.00) | .04 (.00) | .06 (.01) | .08 (.01) |
| COLOR I-Z | .04 (.00) | .09 (.01) | .10 (.01) | .17 (.01) |

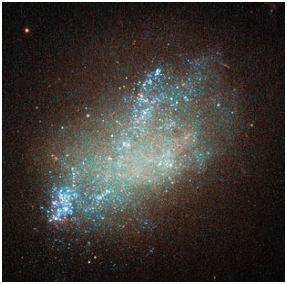

*Figure 6.* An irregular galaxy from the constellation Leo. Because the brightest areas of this galaxy are not near its center, Celeste may misfit this galaxy.
credit: NASA

### 4.3. Model misfit

On real data, galaxy model misfit may have constrained Celeste's performance across the board. Celeste's two-component galaxy model is based on the successful models from Lang & Hogg (2015) and Miller et al. (2013). However, Celeste is a fully generative model, so any misfit between the modeled galaxy and the actual galaxy must be explained by Poisson randomness. The Poisson process is well-suited to modeling the variation in photon count when the underlying rate is correct. But, for galaxies, the rate itself may deviate greatly from the model. With this type of model misfit, the galaxy shape that optimizes the ELBO may be unduly influenced by irregularities in the underlying rate. Figure 6 illustrates a galaxy that would be difficult to fit with any simple parametric model, even if sampled at low spatial resolution. Enhancing Celeste's galaxy model is a promising direction of ongoing research.

Modeling the galaxy-specific parameters ($\rho_s$, $\varphi_s$, $\sigma_s$, and $\theta_s$) as constants to be learned rather than as random variables with prior distributions likely also worsen performance. In some cases galaxies scales' ($\sigma_s$) were much too large; the optimizer may have been using a galaxy to explain a background noise rate that exceeded $\epsilon_{nb}$. These cases likely explain why Primary outperformed Celeste in determining brightness and scale. Though the underlying issue is model misfit (as Celeste's good performance on synthetic data suggests), constraining $\sigma_s$ with a prior distribution could mitigate this effect. Also, a model that assigns more degrees of freedom to galaxies—unconstrained by prior distributions—than stars biases the classification in favor of galaxies. Treating all unknown quantities in Celeste as random is the subject of ongoing research.

## Acknowledgments

This work was supported by the Director, Office of Science, Office of Advanced Scientific Computing Research, Applied Mathematics program of the U.S. Department of Energy under Contract No. DE-AC02-05CH11231.

We thank the anonymous referees for their helpful comments.